\documentclass{article}
\usepackage{rotating}
\usepackage{url}
\usepackage{subcaption}
\usepackage{hyperref}
\usepackage{placeins}
\usepackage{lmodern}
\usepackage{slantsc}

\title{\bfseries Staging Human-computer Dialogs:\\
An Application of\\the Futamura Projections}

\author{{\bfseries Brandon M. Williams} and {\bfseries Saverio Perugini}\\
\normalsize Department of Computer Science\\
\normalsize University of Dayton\\
\normalsize 300 College Park\\
\normalsize Dayton, Ohio\ \ 45469--2160\ \ USA\\
\normalsize Tel: +001 (937) 229--4079, Fax: +001 (937) 229--2193\\
\normalsize E-mail: \url{BrandonWilliamsCS@gmail.com},~\url{saverio@udayton.edu}\\
\normalsize WWW: \url{http://academic.udayton.edu/SaverioPerugini}}

\newcommand{\mix}{mix}
\newcommand{\funceval}[1]{[\![#1]\!]}
\newcommand{\pattident}[1]{#1} % use for pattern identifiers in a math context (no change)
\newcommand{\instident}[1]{\mathtt{#1}} % use for instance identifiers in a math context (teletype font)
\begin{document}
\sloppy

\maketitle
\thispagestyle{empty}

\begin{abstract} We demonstrate an application of the Futamura Projections to
human-computer interaction, and particularly to staging human-computer dialogs.
Specifically, by providing staging analogs to the classical Futamura Projections,
we demonstrate that the Futamura Projections can be applied to the
\textit{staging} of human-computer dialogs in addition to the
\textit{execution} of programs.
\end{abstract}


\section{Introduction}

The \textit{Futamura Projections} are a series of program signatures reported
by~\cite{partialEvaluationComputationProcess} (a reprinting
of~\cite{partialEvaluationComputationProcessOrig}) designed to create a program
that generates compilers by repeated applications of a
\textit{partial evaluator} that iteratively abstracts away aspects of the program
execution process. A partial evaluator transforms a program given any subset of
its input to produce a version of the program that has been specialized to that
input. We use the symbol \texttt{mix} from~\cite{introPartialEvaluation} to
denote the partial evaluation operation because partial evaluation involves a
\texttt{mix}ture of interpretation and code generation.

In this article, we introduce
a model for staging human-computer dialogs based on The Futamura Projections.
In other words, the Futamura Projections provide a way to generate programs
that stage interactions between two participants engaged in a human-computer
dialog for a variety of dialog representations.  Table~\ref{tab:StagingLegend}
is a legend mapping terms and symbols used in this article to their
description.
For a general introduction to the
Futamura Projections, using the same diagramming conventions as this article,
we refer the reader to~\cite{FutamuraTAAI}.

\section{Staging Human-computer Dialogs}
\label{sec:Staging}

\begin{table}
\centering
\caption{Legend of symbols and terms used in
$\S$~\ref{sec:Staging}~and~$\S$\ref{sec:FutamuraStaging}.}
\resizebox{\textwidth}{!}{
\begin{tabular}{|l|l|}
\hline
\multicolumn{1}{|c|}{\textbf{Symbol}} & \multicolumn{1}{c|}{\textbf{Description}} \\
\hline
$\pattident{dialog}$ & A dialog between two parties. \\
$\pattident{p}_{\pattident{n}}$ & A prompt for some input. \\
$\pattident{a}_{\pattident{n}}$ & A response to a prompt. \\
$\pattident{dialog}\ \pattident{success}$ & The result of a successful dialog. \\
$\pattident{DDSL}$ & A domain-specific language designed for representing human-computer-dialogs. \\
$\instident{coffee}\ \instident{dialog}$ & A dialog facilitating a coffee order. \\
$\instident{coffee}\ \instident{as}\ \instident{ordered}$ & Coffee made according to the dialog prompts and responses. \\
$\pattident{DDSL}\ \pattident{compiler}$ & A program that generates stagers when given a dialog specification. \\
$\pattident{dialog}$ & An instance of a dialog specification \textsc{ddsl}. \\
$\pattident{stager}$ & A program that stages a dialog. \\
$\instident{coffee}\ \instident{stager}$ & A stager for the coffee dialog. \\
$\pattident{DDSL}\ \pattident{interpreter}$ & A program that stages a dialog given its specification and responses. \\
$\pattident{DDSL}\ \pattident{compiler}\ \pattident{generator}$ & A program that creates \textsc{ddsl} compilers when given a \textsc{ddsl} interpreter. \\
\hline
\end{tabular}}
\label{tab:StagingLegend}
\end{table}

\begin{figure}[t]
\centering
\resizebox{\textwidth}{!}{
\begin{subfigure}{0.49\textwidth}
	\centering
	\includegraphics[scale=0.70]{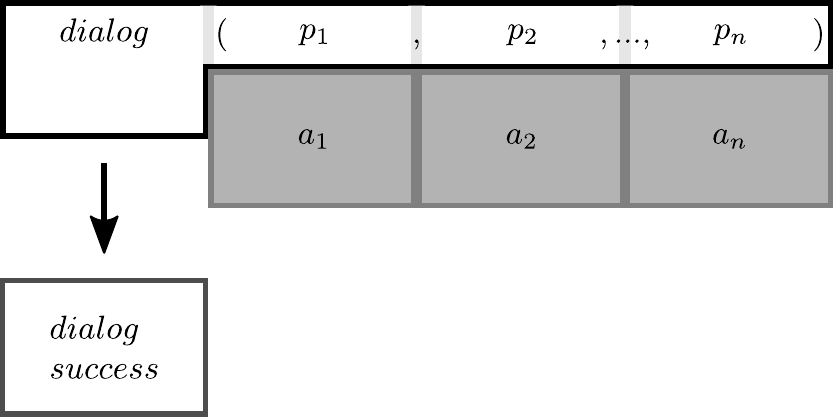}
    \caption{Dialog interaction.}
    \label{fig:DialogPattern}
\end{subfigure}
\begin{subfigure}{0.49\textwidth}
	\centering
	\includegraphics[scale=0.70]{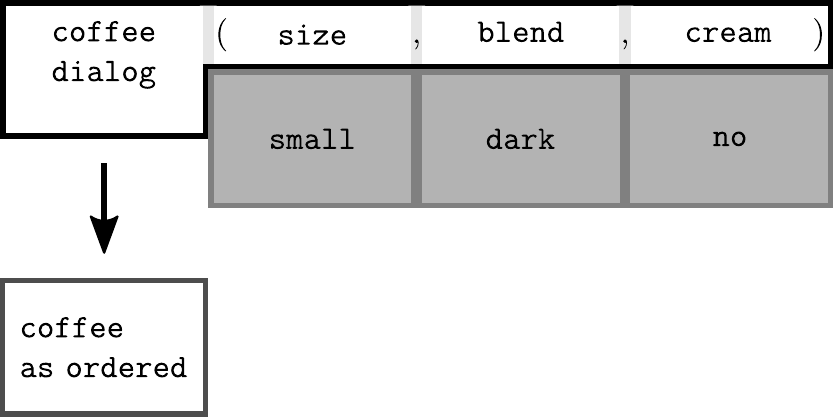}
	\caption{Instance of dialog interaction.}
	\label{fig:DialogExample}
\end{subfigure}}
\caption{Dialog interaction and example instance.}
\end{figure}

A \textit{dialog}, for the purposes of this paper, is a series of interactions
between a user and a computer system, which do not necessarily occur through a
verbal modality. For instance, a user of an installation wizard for an
application program participates in a human-computer dialog~\cite{MIIbook}.
Additionally, dialogs in this paper are not specific to their responses, but represent an ordering
of prompts and set of appropriate responses for a particular purpose of interaction. In
this way, a dialog is analogous to a program: a definition of behavior with results that differ
 based on variable information provided as input. We can represent dialogs
diagrammatically as seen in
Fig.~\ref{fig:DialogPattern} or equationally as
$\funceval{\pattident{dialog}}[\pattident{a_{1}}, \pattident{a_{2}}, ...,
\pattident{a_{n}}] = [\pattident{dialog\ success}]$.  For example, a dialog for
ordering coffee may prompt a customer for the size and bean blend of the
coffee, as well as whether or not to leave room for cream. Such an example
instance is depicted in Fig.~\ref{fig:DialogExample} and described equationally
as $\funceval{\instident{coffee\ dialog}}[\instident{small}, \instident{dark},
\instident{no}] = [\instident{coffee\ as\ ordered}]$.

Although dialogs are analogous to programs, a dialog is not a program and cannot
execute on a computer. Instead,
it must be \textit{staged}\footnote{We are not referring to \textit{staging}
    as the `language construct that allows a program at one stage of evaluation to
    manipulate and specialize a program to be executed at a later
    stage'~\cite{stagingMLHOSC}.  Rather, we are using the term to refer to `providing a platform for
    the progressive interaction of a human-computer dialog.'}
with the assistance of a special program. A program
that structures the interaction of a particular dialog is called a \textit{stager} for that dialog. If
a dialog is analogous to a program, a stager is analogous to a second program, semantically
equivalent to the first, that has been implemented in a natively executable language.
In other words, a stager is analogous to a compiled dialog. The details of staging
human-computer dialogs, and particularly mixed-initiative
dialogs~\cite{WhatIsMII}, are given in~\cite{eics2016}.

\subsection{Dialog DSL Interpretation}

\begin{figure}
\centering
\begin{subfigure}{\textwidth}
	\centering
	\includegraphics[scale=0.75]{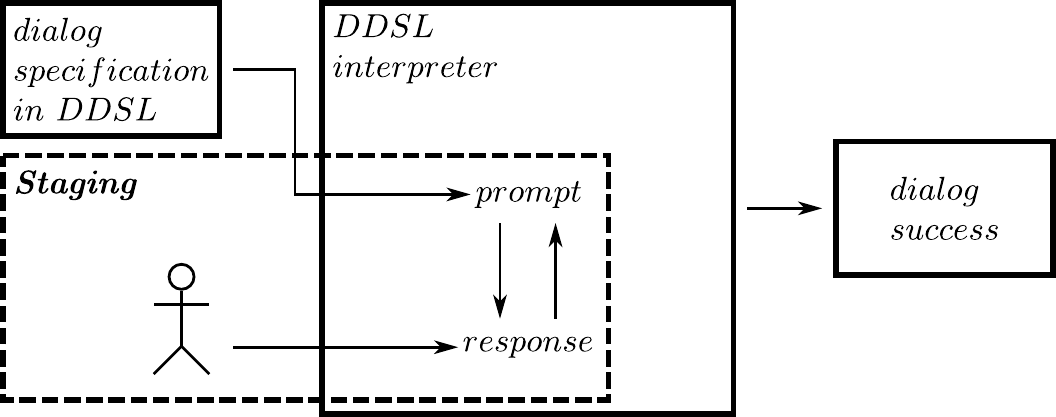}
	\caption{\textsc{ddsl} interpreter operation.}
	\label{fig:DSLInterpreter}
\end{subfigure}
\begin{subfigure}{\textwidth}
	\centering
	\includegraphics[scale=0.75]{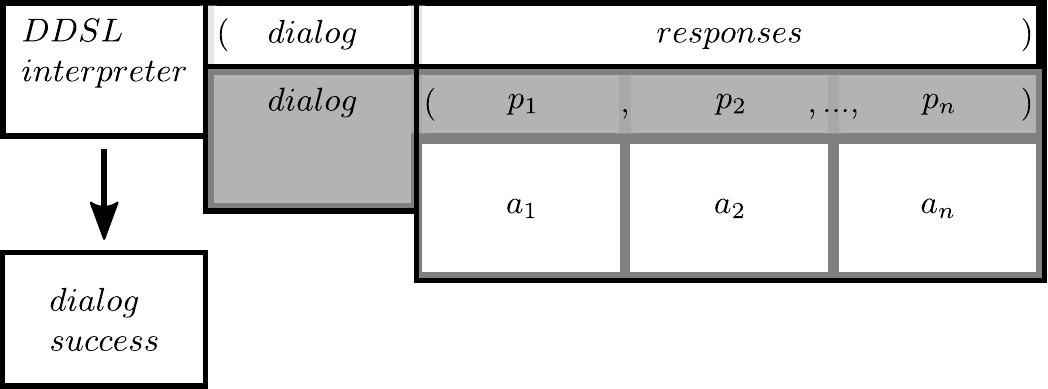}
	\caption{\textsc{ddsl} interpretation process.}
	\label{fig:DSLInterpreterPattern}
\end{subfigure}
\begin{subfigure}{\textwidth}
	\centering
	\includegraphics[scale=0.75]{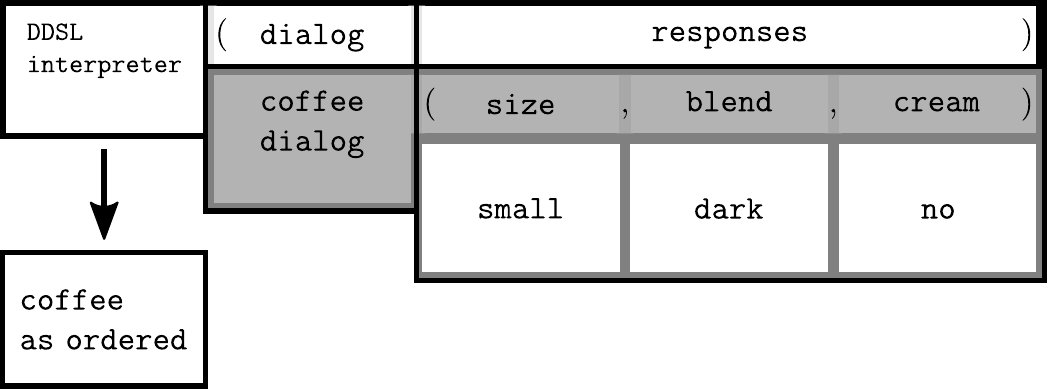}
	\caption{\textsc{ddsl} interpretation example instance.}
	\label{fig:DSLInterpreterExample}
\end{subfigure}
\caption{Staging of dialogs with a \textsc{ddsl} interpreter.}
\end{figure}

Just as high-level programming languages are suited to describing programs to humans,
we can design a special domain-specific language (\textsc{dsl}) suited to describing
dialogs to humans. As with programming languages, we can then write a dialog
\textsc{dsl} (\textsc{ddsl}) interpreter that accepts a dialog specification in that
language and stages it, a process detailed in Fig.~\ref{fig:DSLInterpreter}.
Fig.~\ref{fig:DSLInterpreterPattern} illustrates \textsc{ddsl}
interpretation with our diagram conventions
while the equational representation is
$\funceval{\pattident{DDSL\ interpreter}}[\pattident{dialog}, \pattident{responses}] = [\pattident{dialog\ success}]$.
Fig.~\ref{fig:DSLInterpreterExample} and the expression
$\funceval{\instident{DDSL\ interpreter}}[\instident{coffee\ dialog}, \instident{small}, \instident{dark}, \instident{no}] = [\instident{coffee\ as\ ordered}]$
show the \textsc{ddsl} interpreter staging the coffee dialog from Fig.~\ref{fig:DialogExample}.

\subsection{Dialog DSL Compilation}

\begin{figure}
\centering
\begin{subfigure}{\textwidth}
	\centering
	\includegraphics[scale=0.75]{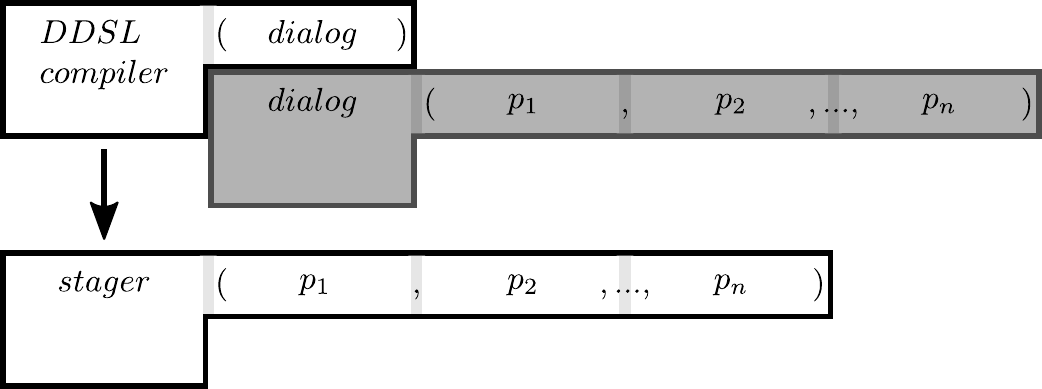}
	\caption{The \textsc{ddsl} compilation process.}
	\label{fig:StagerGeneratorPattern}
\end{subfigure}
\begin{subfigure}{\textwidth}
	\centering
	\includegraphics[scale=0.75]{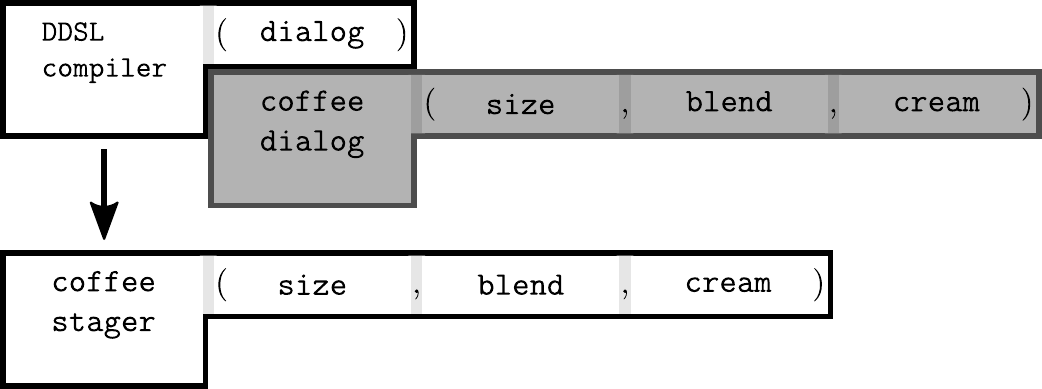}
	\caption{\textsc{ddsl} compilation example instance.}
	\label{fig:StagerGeneratorExample}
\end{subfigure}
\begin{subfigure}{\textwidth}
	\centering
	\includegraphics[scale=0.75]{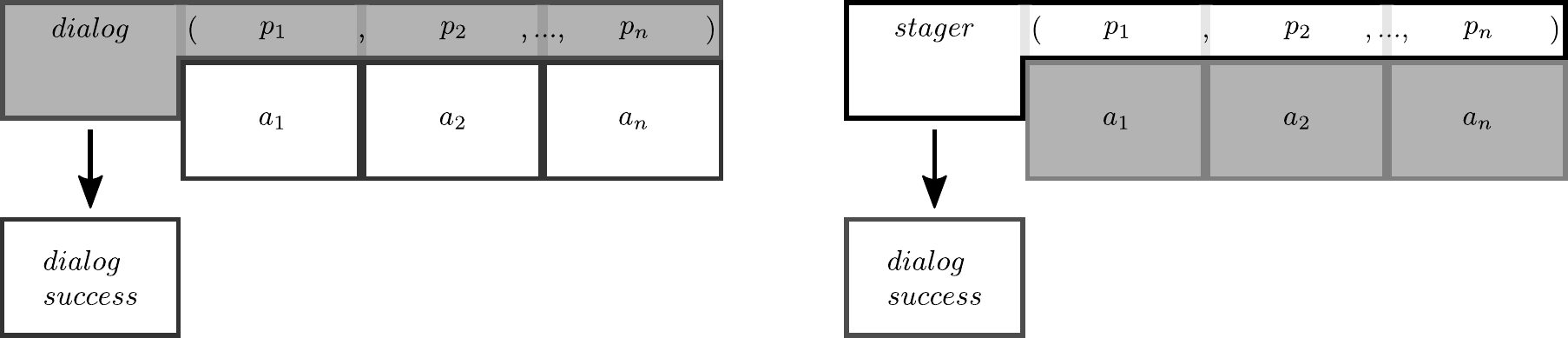}
	\caption{Comparison of \textsc{ddsl} compiler input and output.}
	\label{fig:StagerGeneratorPatternOutput}
\end{subfigure}
\begin{subfigure}{\textwidth}
	\centering
	\includegraphics[scale=0.75]{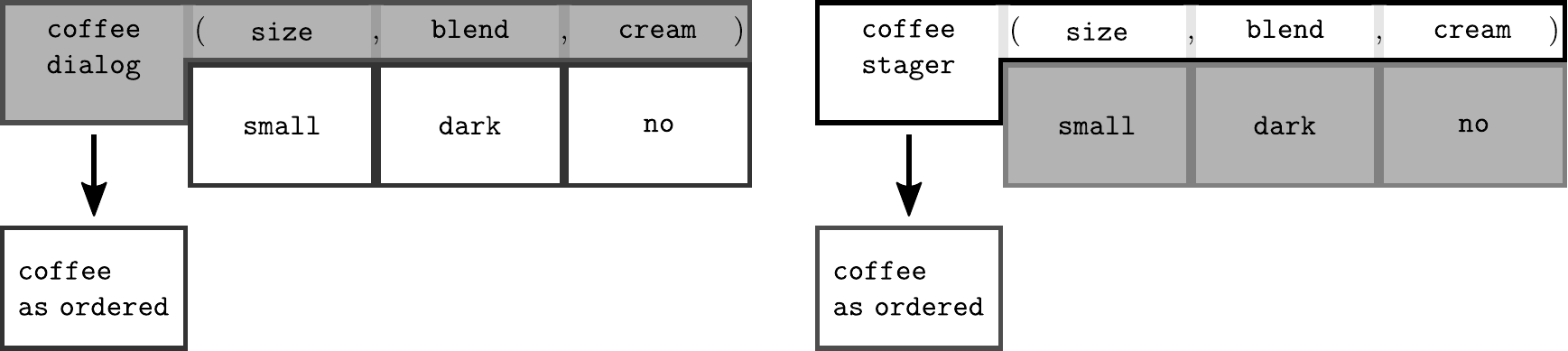}
	\caption{Comparison of \textsc{ddsl} compiler example instance input and output.}
	\label{fig:StagerGeneratorExampleOutput}
\end{subfigure}
\caption{Compilation of \textsc{ddsl} into a stager.}
\end{figure}

Compilers translate programs in a source language to equivalent programs in a target language.
Analogously, we have created a program that takes dialog specifications in a \textsc{ddsl} and
generates a stager for that dialog. We call this stager generator program a
\textsl{\textsc{ddsl} compiler}. A \textsc{ddsl} compiler program is depicted in
Fig.~\ref{fig:StagerGeneratorPattern} and expressed equationally as
$\funceval{\pattident{DDSL\ compiler}}[\pattident{dialog}] = [\pattident{dialog\ stager}]$.
Although a dialog specification is not executable, Fig.~\ref{fig:StagerGeneratorPatternOutput}
demonstrates the behavioral equivalence of the stager and the specification.
The generation of a stager for the coffee dialog is shown in Fig.~\ref{fig:StagerGeneratorExample} and expressed equationally as
$\funceval{\instident{DDSL\ compiler}}[\instident{coffee\ dialog}] = [\instident{coffee\ stager}]$,
and its input and output are compared in Fig.~\ref{fig:StagerGeneratorExampleOutput}.

\section{A Programming Model for Staging Dialogs}
\label{sec:FutamuraStaging}

The Third Futamura Projection produces a program that, given an interpreter for
\textit{any} source programming language, can generate a compiler for that
language. Furthermore, this interpreter need not be thought of exclusively in
the traditional sense as a program interpreter. In particular, it can be
\textit{any} program that conforms to the interpretation signature by accepting
two inputs, namely some source that describes a behavior and the input to that
source, and enacting the behavior. Providing $\instident{mix}$ with a
\textsc{ddsl} interpreter and dialog specifications yields programs meaningful
to staging. Therefore, the Futamura Projections can be applied to the
\textit{staging} of human-computer dialogs in addition to the \textit{execution}
of programs, and this is the primary contribution of this paper.  To facilitate the
staging of dialogs, we provide staging analogs to the classical Futamura
Projections.

\subsection{First Futamura Staging Projection:\\Dialog DSL Compilation}
\begin{figure}
\centering
\begin{subfigure}{\textwidth}
	\centering
	\includegraphics[scale=0.75]{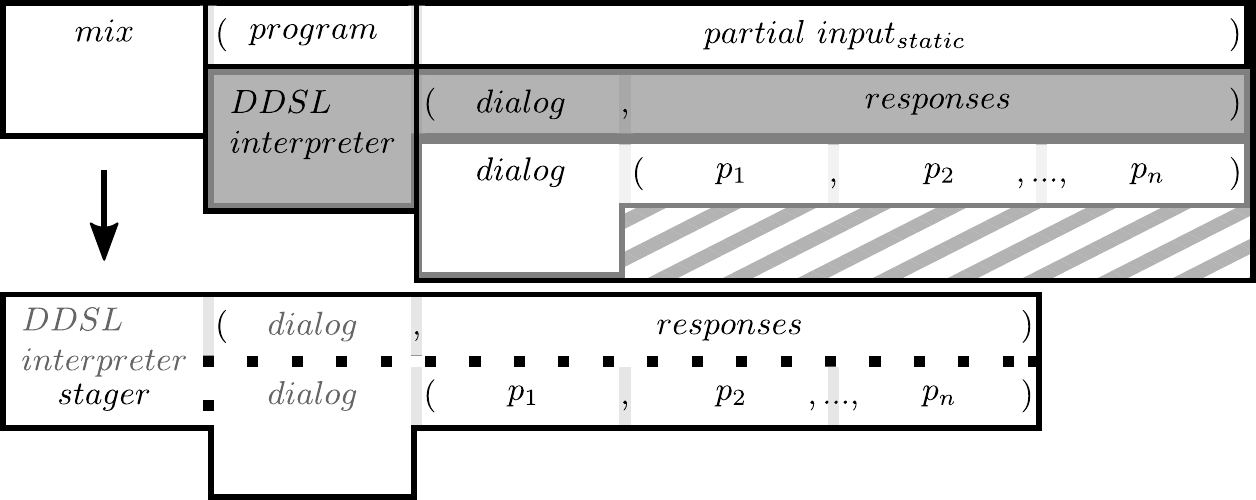}
	\caption{The First Futamura Staging Projection.}
	\label{fig:StagingP1Pattern}
\end{subfigure}
\begin{subfigure}{\textwidth}
	\centering
	\includegraphics[scale=0.75]{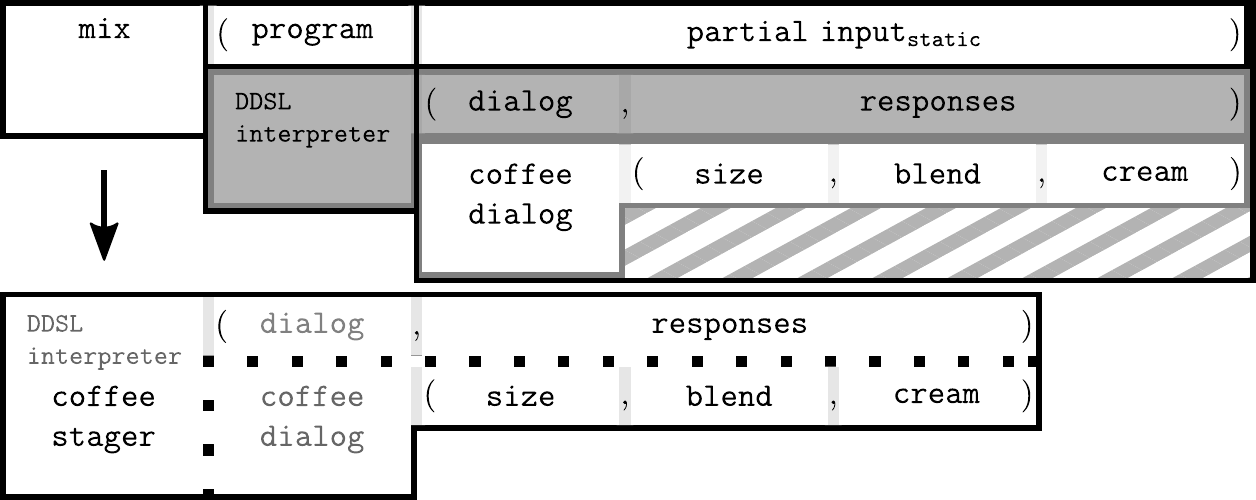}
	\caption{An instance of the First Futamura Staging Projection.}
	\label{fig:StagingP1Example}
\end{subfigure}
\begin{subfigure}{0.62\textwidth}
	\centering
	\includegraphics[scale=0.75]{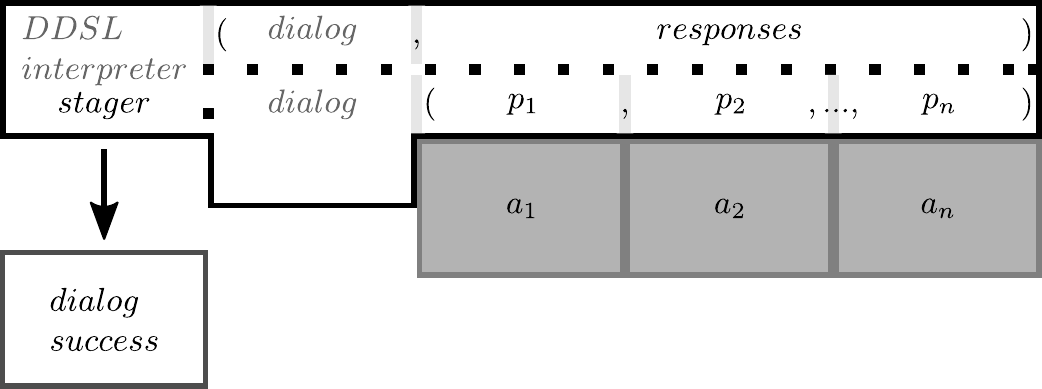}
	\caption{The output of the First Futamura Staging Projection.}
	\label{fig:StagingP1PatternOutput}
\end{subfigure}
\begin{subfigure}{0.6\textwidth}
	\centering
	\includegraphics[scale=0.75]{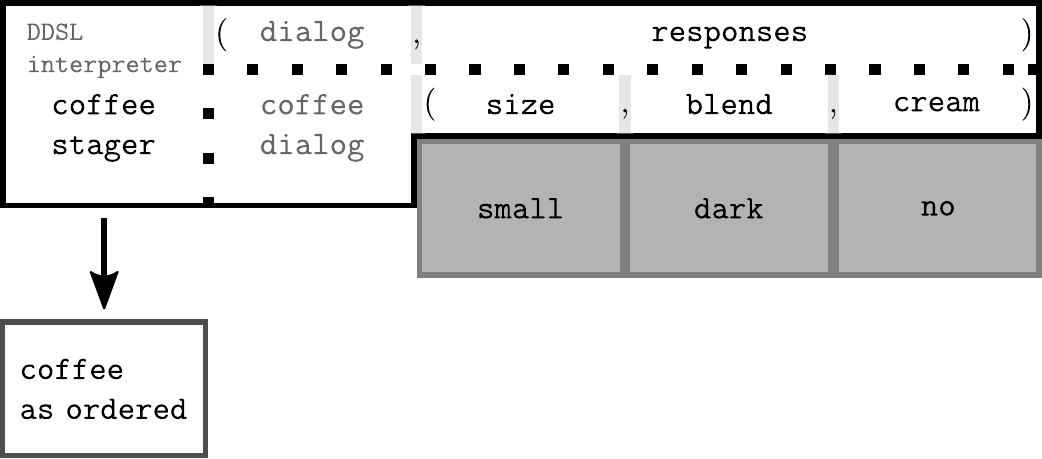}
	\caption{The output of the First Futamura Staging Projection instance.}
	\label{fig:StagingP1ExampleOutput}
\end{subfigure}
\caption{The First Futamura Staging Projection and example instance.}
\end{figure}

Let us first look at the partial evaluation of a \textsc{ddsl} interpreter with a dialog specification as
static input, as shown in Fig.~\ref{fig:StagingP1Pattern} and expressed equationally as
$\funceval{\pattident{\mix{}}}[\pattident{DDSL\ interpreter}, \pattident{dialog}] = [\pattident{stager}]$.
Here $\instident{mix}$ specializes the \textsc{ddsl} interpreter to the dialog, leaving the responses as dynamic input.
The resulting program, shown in Fig.~\ref{fig:StagingP1PatternOutput} and represented equationally as
$\funceval{\pattident{stager}}[\pattident{a_{1}}, \pattident{a_{2}}, ..., \pattident{a_{n}}] = [\pattident{dialog\ success}]$,
accepts the responses and completes the staging of the dialog.
\textit{In other words, the specialized program is a stager for the input dialog}.
The partial evaluator has effectively generated a stager from a \textsc{ddsl} dialog specification.
This transformation is analogous to the compilation of a source program into a
target program in the First Futamura Projection pattern.
Applying the pattern to the coffee dialog instance, Fig.~\ref{fig:StagingP1Example} depicts the partial evaluation of a \textsc{ddsl} interpreter with the coffee specification as input. This process can be expressed equationally as
$\funceval{\instident{\mix{}}}[\instident{DDSL\ interpreter}, \instident{coffee\ dialog}] = [\instident{coffee\ stager}]$.
The resulting program, examined in Fig.~\ref{fig:StagingP1ExampleOutput} and expressed equationally as
$\funceval{\instident{coffee\ stager}}[\instident{small}, \instident{dark}, \instident{no}] = [\instident{coffee\ as\ ordered}]$,
stages the coffee dialog with any responses to produce the result (i.e., the ordered coffee).

\begin{quote}
\textbf{First Futamura Staging Projection}: A partial evaluator, with the aid
of a \textsc{ddsl} interpreter, can generate stagers.
\end{quote}

\subsection{Second Futamura Staging Projection:\\DDSL Compiler Generation}

\begin{figure}
\centering
\begin{subfigure}{\textwidth}
	\centering
	\includegraphics[scale=0.75]{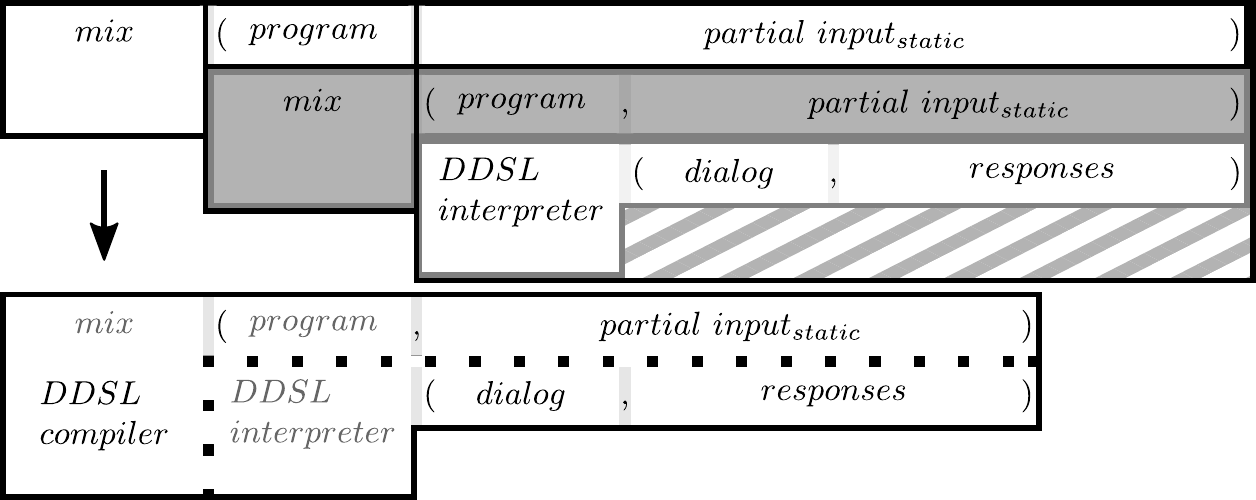}
	\caption{The Second Futamura Staging Projection.}
	\label{fig:StagingP2Pattern}
\end{subfigure}
\begin{subfigure}{\textwidth}
	\centering
	\includegraphics[scale=0.75]{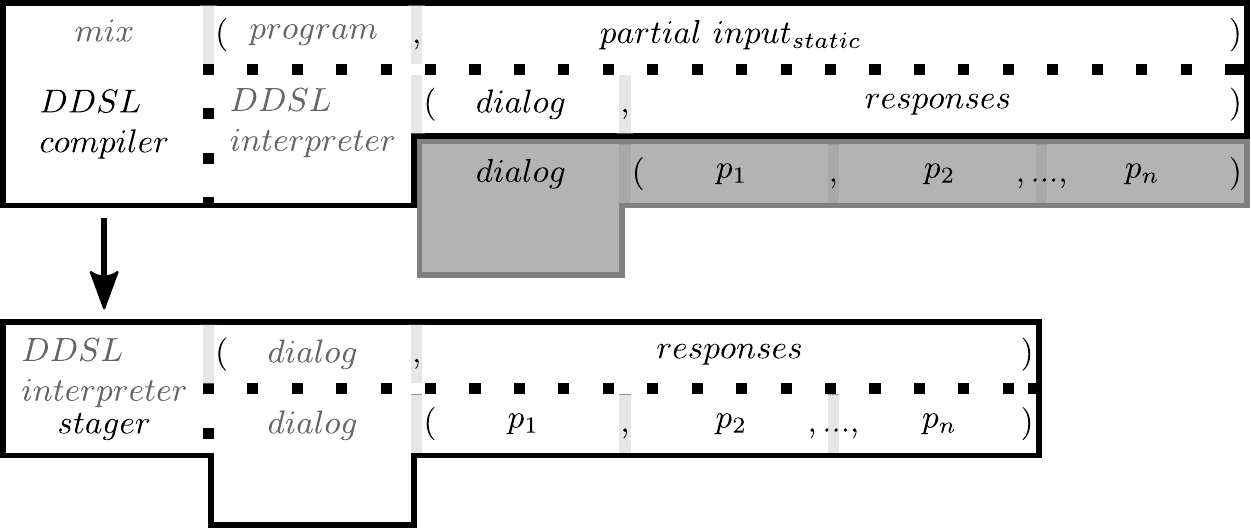}
	\caption{The output of the Second Futamura Staging Projection.}
	\label{fig:StagingP2PatternOutput}
\end{subfigure}
\begin{subfigure}{\textwidth}
	\centering
	\includegraphics[scale=0.75]{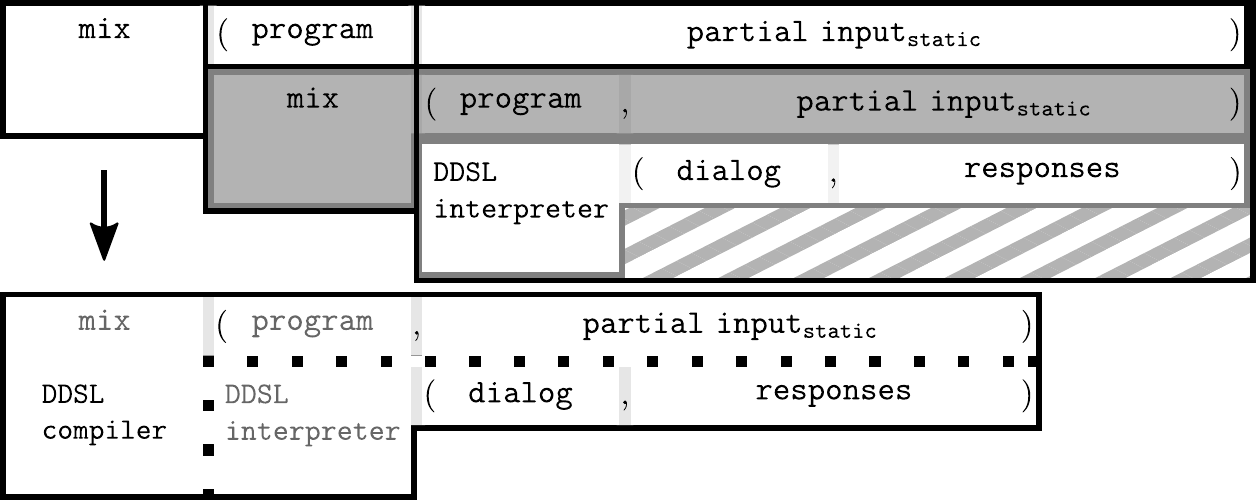}
	\caption{An instance of the Second Futamura Staging Projection.}
	\label{fig:StagingP2Example}
\end{subfigure}
\begin{subfigure}{\textwidth}
	\centering
	\includegraphics[scale=0.75]{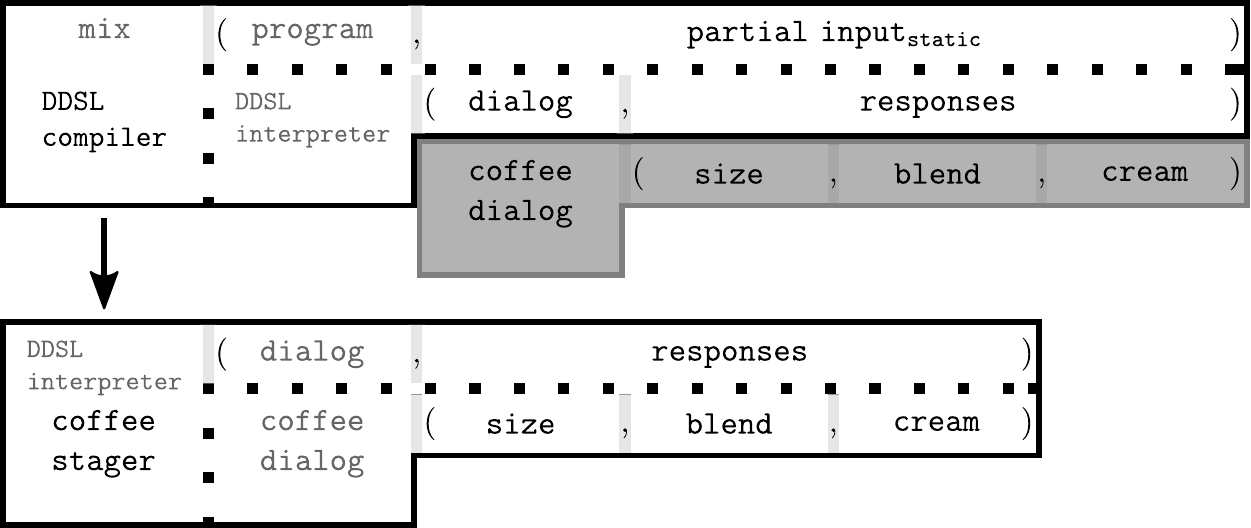}
	\caption{The output of the Second Futamura Staging Projection instance.}
	\label{fig:StagingP2ExampleOutput}
\end{subfigure}
\caption{The Second Futamura Staging Projection and example instance.}
\end{figure}

We have applied the pattern of the First Futamura Projection to dialog staging to produce the
First Futamura Staging Projection. Because the Second Futamura Projection is just a partial
evaluation of the first, we can apply the pattern of the Second Futamura Projection to staging by
partially evaluating the First Futamura Staging Projection. This Projection is shown in
Fig.~\ref{fig:StagingP2Pattern} and expressed as
$\funceval{\pattident{\mix{}}}[\pattident{\mix{}}, \pattident{DDSL\ interpreter}] = [\pattident{DDSL\ compiler}]$,
where $\instident{mix}$ itself is being partially evaluated with the \textsc{ddsl} interpreter
as static input.
\textit{The resulting program, as with the result of the Second Futamura Projection, captures the
behavior of the previous Projection.}
Fig.~\ref{fig:StagingP2PatternOutput}
(represented equationally as $\funceval{\pattident{DDSL\ compiler}}[\pattident{dialog}] = [\pattident{dialog\ stager}]$)
demonstrates this by accepting a dialog specification and producing a stager for it. Notice that it has
the same shape and labels as Fig.~\ref{fig:StagingP1Pattern}.
The diagrams are the same as those for the classical Futamura Projections, but with the source
language interpreter replaced with a \textsc{ddsl} interpreter and the source
and target programs replaced with a dialog specification and stager, respectively. Instead of
abstracting the source program away from a $\instident{mix}$-based program compilation process,
we are abstracting the dialog specification away from a $\instident{mix}$-based \textsc{ddsl} compilation process.

The Second Futamura Staging Projection is applied to the coffee dialog in Figs.~\ref{fig:StagingP2Example} and \ref{fig:StagingP2ExampleOutput}. In the first, $\instident{mix}$ is specialized to the \textsc{ddsl} interpreter for the dialog. In the second, the output program completes the partial evaluation of the \textsc{ddsl} interpreter and produces a stager for the coffee dialog. These figures are represented equationally as
$\funceval{\instident{\mix{}}}[\instident{\mix{}}, \instident{DDSL\ interpreter}] = [\instident{DDSL\ compiler}]$
and
$\funceval{\instident{DDSL\ compiler}}[\instident{coffee\ dialog}] = [\instident{coffee\ stager}]$,
respectively.

\begin{quote} \textbf{Second Futamura Staging Projection}: A partial evaluator,
by making use of another instance of itself and a \textsc{ddsl} interpreter, can generate
\textsc{ddsl} compilers.  \end{quote}

\subsection{Third Futamura Staging Projection:\\Generation of DDSL Compiler Generators}

\begin{figure}
\centering
\begin{subfigure}{\textwidth}
	\centering
	\includegraphics[scale=0.75]{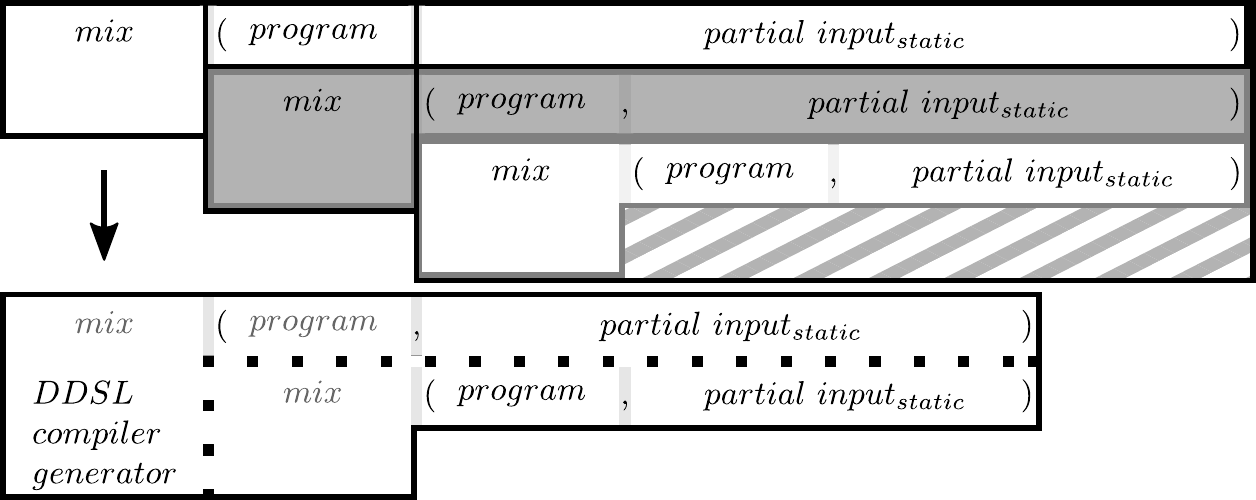}
	\caption{The Third Futamura Staging Projection.}
	\label{fig:StagingP3Pattern}
\end{subfigure}
\begin{subfigure}{\textwidth}
	\centering
	\includegraphics[scale=0.75]{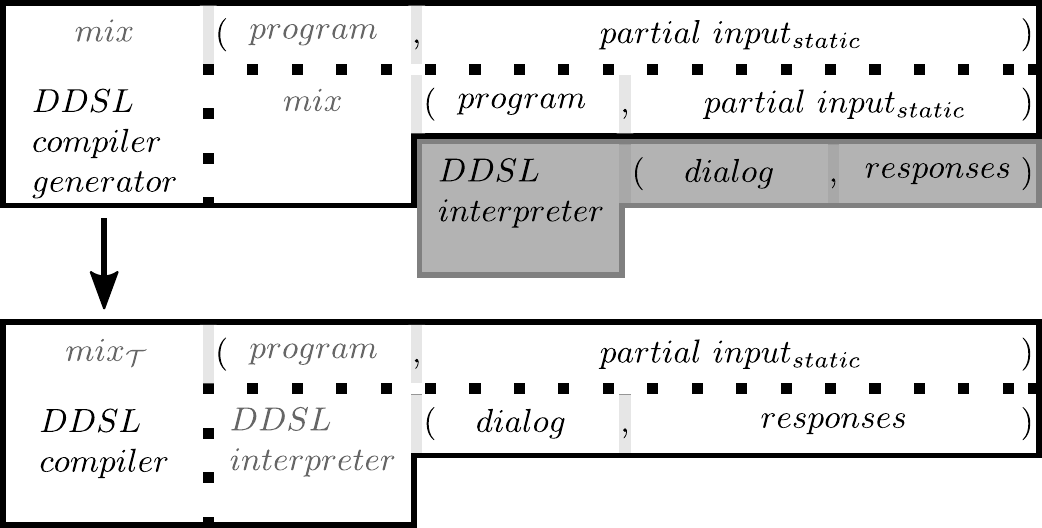}
	\caption{The output of the Third Futamura Staging Projection.}
	\label{fig:StagingP3PatternOutput}
\end{subfigure}
\caption{The Third Futamura Staging Projection.}
\end{figure}

Now we abstract the staging process one final degree by partially performing the Second Futamura Staging Projection
while leaving the \textsc{ddsl} interpreter as dynamic input.
Fig.~\ref{fig:StagingP3Pattern} shows this third Projection, which is expressed equationally as 
$\funceval{\pattident{\mix{}}}[\pattident{\mix{}}, \pattident{\mix{}}] = [\pattident{DDSL\ compiler\ generator}]$.
Aside from the label that the result is a generator of \textsl{\textsc{ddsl}} compilers, there is nothing specific to
dialogs or staging in this projection.
\textit{Because there is no mention of a generic input program, a dialog specification, or either variety of
interpreter (i.e., program or staging), the Third Futamura Staging Projection is the same as
the Third Futamura Projection}.
However, when given different varieties of interpreter as input, the generator program serves different roles.
The result of the Third Futamura Projection generates a traditional compiler when given a programming language
interpreter, but in the context of the Futamura Staging Projections, it generates a \textsc{ddsl} compiler from
a \textsc{ddsl} interpreter.
The latter is depicted in Fig.~\ref{fig:StagingP3PatternOutput} and expressed equationally as 
$\funceval{\pattident{DDSL\ compiler\ generator}}[\pattident{DDSL\ interpreter}] = [\pattident{DDSL\ compiler}]$.

\begin{quote} \textbf{Third Futamura Staging Projection}: A partial evaluator,
by making use of two additional instances of itself, can generate a program that generates
\textsc{ddsl} compilers.  \end{quote}

\subsection{Summary: Futamura Staging Projections}

Each human-computer dialog Staging Projection, akin to each Futamura Projection, abstracts away a
previous process by partial evaluation.  The First Futamura Staging Projection
treats the dialog responses as dynamic input. The Second Staging Projection furthers the abstraction, this time
with the dialog specification. The Third Staging Projection allows the
\textsc{dsl} used for expressing the dialog to vary by treating the
\textsc{ddsl} interpreter as dynamic input. The Futamura Staging Projections extend the program-to-dialog analogy
when each source-program-specific element is replaced by its dialog staging analog.
In other words, replacing the source program
with a dialog specification and the interpreter with a \textsc{ddsl} interpreter
in turn replaces the target program with a stager and the compiler with a
\textsc{ddsl} compiler.  As mentioned earlier, the program compiler generator and the \textsc{ddsl} compiler generator
(i.e., the results of third of each series of projections) are not
just analogous, but identical.  Table~\ref{tab:StagingSummary} juxtaposes the
related equations and diagrams from both $\S$~\ref{sec:Staging} and
$\S$~\ref{sec:FutamuraStaging} in each row to make their relationships more
explicit.  Each row of Table~\ref{tab:StagingProjectionSummary} succinctly
summarizes each Staging Projection by associating each side of its equational representation with the
corresponding figure from $\S$~\ref{sec:FutamuraStaging}.

\begin{sidewaystable}
\caption{Juxtaposition of related equations and diagrams from $\S$~\ref{sec:Staging}~and~$\S$~\ref{sec:FutamuraStaging}.}
\resizebox{\textwidth}{!}{
\begin{tabular}{|l|l|l|l|l|l|}
\hline
\multicolumn{1}{|c|}{\textbf{Fig.}} & \multicolumn{1}{c|}{\textbf{Equational Notation}}
	& \multicolumn{1}{c|}{\textbf{Input Description}} & \multicolumn{1}{c|}{\textbf{Output Description}}
	& \multicolumn{1}{c|}{\textbf{Similar Fig.}} & \multicolumn{1}{c|}{\textbf{Output Fig.}} \\
\hline
\ref{fig:DialogPattern} & $\funceval{\pattident{dialog}}[\pattident{a_{1}}, \pattident{a_{2}}, ..., \pattident{a_{n}}] = [\pattident{dialog\ success}]$
	& A list of responses for the dialog. & The results of a successfully completed dialog. & \multicolumn{1}{c|}{N/A} & \multicolumn{1}{c|}{N/A} \\
\hline
\ref{fig:DSLInterpreterPattern}
	& $\funceval{\pattident{DDSL\ interpreter}}[\pattident{dialog}, \pattident{responses}] = [\pattident{dialog\ success}]$
	& A dialog specification in the \textsc{ddsl} along with its responses. & The results of a successfully completed dialog. & \multicolumn{1}{c|}{N/A} & \multicolumn{1}{c|}{N/A} \\
\hline
\ref{fig:StagerGeneratorPattern}
	& $\funceval{\pattident{DDSL\ compiler}}[\pattident{dialog}] = [\pattident{dialog\ stager}]$
	& A dialog specification in the \textsc{ddsl}. & A program that stages the dialog. & \multicolumn{1}{c|}{N/A} & \multicolumn{1}{c|}{\ref{fig:StagerGeneratorPatternOutput}} \\
\hline

\ref{fig:StagingP1Pattern}
	& $\funceval{\pattident{\mix{}}}[\pattident{DDSL\ interpreter}, \pattident{dialog}]
		 = [\pattident{stager}]$
	& An interpreter for the \textsc{ddsl} and a dialog specification in the \textsc{ddsl}. & A program that stages the dialog. & \multicolumn{1}{c|}{N/A} & \multicolumn{1}{c|}{\ref{fig:StagingP1PatternOutput}} \\
\hline
\ref{fig:StagingP1PatternOutput}
	& $\funceval{\pattident{stager}}[\pattident{a_{1}}, \pattident{a_{2}}, ..., \pattident{a_{n}}]
		 = [\pattident{dialog\ success}]$
	&  A list of responses for the dialog. & The results of a successfully completed dialog. & \multicolumn{1}{c|}{\ref{fig:DSLInterpreterPattern}} & \multicolumn{1}{c|}{N/A} \\
\hline

\ref{fig:StagingP2Pattern}
	& $\funceval{\pattident{\mix{}}}[\pattident{\mix{}}, \pattident{DDSL\ interpreter}]
		 = [\pattident{DDSL\ compiler}]$
	& $\pattident{\mix{}}$ and an interpreter for the \textsc{ddsl}. & A \textsc{ddsl} compiler.
	& \multicolumn{1}{c|}{N/A} & \multicolumn{1}{c|}{\ref{fig:StagingP2PatternOutput}} \\
\hline
\ref{fig:StagingP2PatternOutput}
	& $\funceval{\pattident{DDSL\ compiler}}[\pattident{dialog}]
		 = [\pattident{dialog\ stager}]$
	& A dialog specification in the \textsc{ddsl}. & A program that stages the dialog.
	& \multicolumn{1}{c|}{\ref{fig:StagingP1Pattern}} & \multicolumn{1}{c|}{\ref{fig:StagingP1PatternOutput}} \\
\hline

\ref{fig:StagingP3Pattern}
	& $\funceval{\pattident{\mix{}}}[\pattident{\mix{}}, \pattident{\mix{}}]
		 = [\pattident{DDSL\ compiler\ generator}]$
	& Two instances of $\pattident{\mix{}}$. & A program that generates \textsc{ddsl} compiler & \multicolumn{1}{c|}{N/A} & \multicolumn{1}{c|}{\ref{fig:StagingP3PatternOutput}} \\
\hline
\ref{fig:StagingP3PatternOutput}
	& $\funceval{\pattident{DDSL\ compiler\ generator}}[\pattident{DDSL\ interpreter}]
		 = [\pattident{DDSL\ compiler}]$
	& An interpreter for the \textsc{ddsl}. & A \textsc{ddsl} compiler. & \multicolumn{1}{c|}{\ref{fig:StagingP2Pattern}} & \multicolumn{1}{c|}{\ref{fig:StagingP2PatternOutput}} \\

\hline
\end{tabular}}
\label{tab:StagingSummary}

\caption{Summary of Futamura Staging Projections.}
\resizebox{\textwidth}{!}{
\begin{tabular}{|l|l|l|l|l|l|}
\hline
\multicolumn{1}{|c|}{\textbf{Projection}} & \multicolumn{1}{c|}{\textbf{Description}}
	& \multicolumn{1}{|c|}{\textbf{Equational Notation}}
	& \multicolumn{1}{|c|}{\textbf{Fig.}} & \multicolumn{1}{|c|}{\textbf{ Output Fig.}} \\
\hline

\multicolumn{1}{|c|}{1} & $\instident{mix}$ can generate stagers.
	& $\funceval{\pattident{\mix{}}}[\pattident{DDSL\ interpreter}, \pattident{dialog}]
					= [\pattident{dialog\ stager}]$
	& \ref{fig:StagingP1Pattern} & \multicolumn{1}{c|}{\ref{fig:StagingP1PatternOutput}} \\
\hline

\multicolumn{1}{|c|}{2} & $\instident{mix}$ can generate \textsc{ddsl} compilers.
	& $\funceval{\pattident{\mix{}}}[\pattident{\mix{}}, \pattident{DDSL\ interpreter}]
		 = [\pattident{DDSL\ compiler}]$
	& \ref{fig:StagingP2Pattern} & \multicolumn{1}{c|}{\ref{fig:StagingP2PatternOutput}} \\
\hline

\multicolumn{1}{|c|}{3} & $\instident{mix}$ can generate a program that generates \textsc{ddsl} compilers.
	& $\funceval{\pattident{\mix{}}}[\pattident{\mix{}}, \pattident{\mix{}}]
		 = [\pattident{DDSL\ compiler\ generator}]$
	& \ref{fig:StagingP3Pattern} & \multicolumn{1}{c|}{\ref{fig:StagingP3PatternOutput}} \\

\hline
\end{tabular}}
\label{tab:StagingProjectionSummary}

\hspace{2em}

\end{sidewaystable}

\subsection{Conclusion}

We applied the Futamura Projections to the staging of human-computer dialogs.
Partial evaluation, through the Futamura Staging Projections, can be used to generate stagers,
\textsc{ddsl} compilers, and programs that themselves generate \textsc{ddsl} compilers.
Although the scope of the Futamura Projections has been largely limited to the
programming languages research community, we are optimistic that this article
has provided a programming model for using the Projections in staging
and made a case for their role in building powerful programming abstractions in human-computer interaction.

\end{document}